\documentclass[12pt]{article}
\usepackage[dvips]{color,graphicx}
\DeclareGraphicsExtensions{.eps}

\newcommand{\gs}{\ensuremath{g_s}} 
\newcommand{\ap}{\ensuremath{\alpha'}} 
\newcommand{\ls}{\ensuremath{l_s}} 

\def\p{\partial}

\newcommand{\tr}{\mathop{\rm Tr}}

\newcommand{\cL}{\mathcal{L}}

\newcommand{\cN}{{\mathcal{N}}}

\newcommand{\bS}{{\mathbf{S}}}

\newcommand{\bZ}{{\mathbf{Z}}}

\title{\bf Energy Loss of Gluons, Baryons and\\ $k$-Quarks in an $\cN=4$ SYM Plasma}
\author{Mariano Chernicoff\footnote{e-mail: mariano@nucleares.unam.mx}
~and Alberto G\"uijosa\footnote{e-mail: alberto@nucleares.unam.mx}
\\{\small Departamento de F\'{\i}sica de Altas Energ\'{\i}as,
Instituto de Ciencias Nucleares}\\ {\small Universidad Nacional
Aut\'onoma de M\'exico}\\
{\small Apdo. Postal 70-543, M\'exico D.F. 04510}}
\date{}

\begin{document}
\maketitle

\vspace*{-1cm}

\begin{abstract}
We consider different types of external color sources that move through a strongly-coupled
thermal $\cN=4$ super-Yang-Mills plasma, and calculate, via the AdS/CFT correspondence, the
dissipative force (or equivalently, the rate of energy loss) they experience. A bound state
of $k$ quarks in the totally antisymmetric representation is found to feel a force with a
nontrivial $k$-dependence. Our result for $k$=1 (or $k=N-1$) agrees at large $N$ with the
one obtained recently by Herzog \emph{et al.} and Gubser, but contains in addition an
infinite series of $1/N$ corrections. The baryon ($k=N$) is seen to experience no drag.
Finally, a heavy gluon is found to be subject to a force which at large $N$ is twice as
large as the one experienced by a heavy quark, in accordance with gauge theory expectations.
\end{abstract}

\section{Introduction and Summary}

A considerable effort has recently been invested in the study of strongly-coupled thermal
non-Abelian plasmas by means of the AdS/CFT correspondence \cite{malda,gkpw,magoo}. The main
motivation is the hope of making contact with experimental data on the strongly-coupled
quark-gluon plasma (sQGP) that has been produced at RHIC \cite{rhic} and will be produced at
LHC \cite{alice} (for reviews, see, e.g., \cite{qgprev}). At our present stage of knowledge,
this will be feasible only if real-world QCD can be reasonably well approximated by at least
one of the various `QCD-like' gauge theories whose dual description is known. In the past
few years, encouraging signs in this direction have emerged even for the most rudimentary
example \cite{malda}, $SU(N)$ $\cN=4$ super-Yang-Mills (SYM), which at zero temperature is
completely unlike QCD, but at finite temperature is in various respects analogous to
deconfined QCD. Indeed, the numerical values of several properties of the sQGP appear to be
in the ballpark of the corresponding AdS/CFT predictions for a strongly-coupled SYM plasma,
including its strong-to-weak-coupling entropy ratio \cite{threequarters,karsch} and its
ratio of shear viscosity to entropy density \cite{pss,teaney}.\footnote{Notice, however,
that the putative similarity in viscosity does not hold at \emph{weak} coupling \cite{hjm}.}
These indications have spurred intense research on various fronts, including attempts to
achieve a more realistic model of the sQGP by incorporating the effect of its expansion
and/or its finite extent \cite{nastase,sin,jp,gubseretal4}.

Much of the recent activity in this area has focused on determining the rate at which the
thermal non-Abelian plasma dissipates energy. The drag force experienced by a heavy quark
that ploughs through a strongly-coupled $\cN=4$ SYM plasma was determined in
\cite{hkkky,gubser}.\footnote{The corresponding weakly-coupled calculation was carried out
in \cite{cv}.} The closely related heavy quark diffusion coefficient was computed in
\cite{ct,hkkky}. Previous related work was carried out in \cite{sin1}; generalizations can
be found in \cite{herzog,cacg1,sin2,cacg2,mtw,ntw,talavera}. In interesting followup work,
the authors of \cite{gubseretal1,gxz,gubseretal2} studied the profile of the coherent
gluonic fields set up by the joint quark-plasma system, which are responsible for taking
energy away from the moving quark. Their results appear to be consistent with
phenomenological expectations of `conical flow' \cite{conical}.

An important measure of energy loss used in phenomenological models of medium-induced
radiation (for reviews see \cite{baier}) is the jet-quenching parameter $\hat{q}$, defined
as the average squared transverse momentum transferred to the quark by the medium, per unit
distance travelled. The authors of \cite{liu} suggested that $\hat{q}$ could be identified
with the logarithm of a certain lightlike Wilson loop,\footnote{Later work emphasized that
this lightlike loop can be continuously connected with neighboring \emph{timelike} loops
only if the latter are taken to be traced by a source that is not completely pointlike
\cite{liu2,cgg}. Alternatively, the loop of \cite{liu} may be thought of as a limit of
\emph{spacelike} loops traced by an ordinary pointlike source \cite{cgg}. The physical
significance of either of these statements remains to be understood; we will come back to
the second statement towards the end of Section \ref{kquarksec}.} and then used AdS/CFT to
compute the latter for $\cN=4$ SYM. The authors of \cite{hkkky} argued that a prediction for
$\hat{q}$ in $\cN=4$ SYM could be extracted, under certain assumptions, from their value for
the drag force through use of the Langevin equation. Their result does not agree with that
of \cite{liu}, so there is some controversy on whether the lightlike Wilson loop employed in
the latter work really computes the jet quenching parameter as defined in \cite{baier}. Be
that as it may, a number of subsequent works have applied the prescription of \cite{liu} in
more elaborate contexts \cite{buchel,vp,cacg2,lin,sfetsos,edelstein,ntw}, finding results
whose qualitative form resembles that of corresponding drag forces
\cite{herzog}-\cite{talavera}, even if the detailed functional form of the two sets of
quantities generally disagrees.

Several works have also considered the case where the plasma is probed not with a single
quark but with a quark-antiquark pair. Such mesons were found to feel no drag force
\cite{sonnenschein,liu2,cgg}: being color-neutral, they do not set up the long-range gluonic
fields\footnote{The color field profile set up by a meson was explicitly determined in
\cite{cg} in the zero-temperature case.} that could transport energy away from them. This,
however, is only true as long as the quark and antiquark are bound, which is of course made
difficult by the screening effect of the plasma. The relevant $q$-$\bar{q}$ potential in the
presence of a strongly-coupled $\cN=4$ SYM plasma was computed some time ago in the case
where the pair is static with respect to the plasma \cite{theisen,brandhuber}, and  recently
in the more general case where the pair moves with an arbitrary velocity \cite{cgg}. The
upshot is that the quark and antiquark become unbound, and consequently experience a drag
force, if their separation exceeds a certain screening length whose velocity- and
temperature-dependence was obtained in \cite{liu2,cgg} (see also \cite{sonnenschein}).
Various interesting extensions and refinements of these calculations have been performed in
\cite{elena,argyres,sfetsos2,gubseretal3}. As first emphasized in \cite{liu2}, the results
should have implications for charmonium suppression in the sQGP.

In this paper we take yet another step along this road by considering additional moving
probes of the $\cN=4$ SYM plasma, and using the AdS/CFT toolbox to determine their rates of
energy loss. The probe of primary phenomenological interest is a gluon, because hard partons
traversing the sQGP are expected to be gluons somewhat more often than they are quarks (in
spite of which the hadrons with high transverse momentum detected after the collision come
primarily from quarks \cite{guy}). One of our aims in this paper is therefore to compute the
drag force experienced by a heavy gluon. A sketch of how one may go about doing this using a
string and an antistring was given previously in \cite{gubserpitp} (a related discussion may
be found in \cite{gubseretal3}).

We will model the gluon here as a pointlike external color source in the adjoint
representation. {}From the theoretical perspective, it is also interesting to consider
probes of the plasma that transform in other representations of the $SU(N)$ gauge group. The
technology for describing such sources in the dual AdS language has been developed only
recently, in the context of Wilson loop computations.

The recipe for calculating Wilson loops in the fundamental representation has been known for
a long time \cite{maldawilson,reyee,dgo}, and involves the identification of a string as the
AdS counterpart of a fundamental color source. The case of Wilson loops in higher
representations was discussed qualitatively already in \cite{go} in terms of a collection of
coincident strings, but only began to be understood more systematically after the beautiful
work \cite{df}. It was shown there that a D3-brane that carries electric flux\footnote{The
relevance of D3-branes had been noted already in the original work \cite{reyee}, based on
the Born-Infeld string construction \cite{cm}.} (i.e., a D3-F1 bound state) provides a more
accurate description that correctly captures the effects of the interactions among the
strings, which give rise to an infinite series of non-planar corrections in the gauge theory
side \cite{dg}. This idea was elaborated on in interesting ways in the three simultaneous
works \cite{hpk,yamaguchi,gp}. The last paper, in particular, provided an elegant and
complete dictionary for calculating Wilson loops in an arbitrary representation, using a
collection of \emph{either} D3-branes \emph{or} D5-branes that carry electric flux. This
 prescription was derived explicitly in \cite{gp} for the special case of half-BPS
Wilson loops in zero-temperature $\cN=4$ SYM, but given the motivation of \cite{df} it
should apply much more generally, as has indeed been assumed in subsequent works
\cite{rodriguez,hartnoll,hpk2,ch,grt,ty}. For our purposes, the bottom line is that the
results of \cite{gp,yamaguchi} unambiguously identify D3-branes or D5-branes with
appropriate electric fluxes as the AdS counterparts of the various external color sources
that we wish to push through the SYM plasma.

To start out with, we will employ this identification to compute
the dissipative force experienced by a collection of $k$ quarks in
the totally antisymmetric representation, which according to
\cite{yamaguchi,gp} is dual to a single D5-brane with $k$ units of
electric flux. Arbitrary Wilson loops in this same representation
were calculated recently in the nice paper \cite{hartnoll}, and
the D5-brane we need to consider here is just a particular case of
the ones studied there. We begin  in Section \ref{d5sec} by
setting up the problem and working out the
 D5-brane embedding of interest. In the process, we discover that the
 constant-polar-angle D5-brane configurations (\ref{thetak})-(\ref{esol})
 obtained in \cite{pr,cpr}, which are basic
 building blocks in the dictionary of \cite{yamaguchi,gp}, are in fact incomplete: they are
 missing an `end cap' that lies at the AdS boundary and whose presence is crucial to
 reproduce the correct energy and $SU(4)$ charge of the dual $k$-quark
 system.\footnote{In the extremal case, it is only
 after including the contribution of this end cap that the energy of the system is proportional
 to its charge, resolving the
 puzzle encountered in \cite{pr}.} The complete
 D5-brane embeddings turn out to be a special case of the solutions obtained in the
 earlier work \cite{cgs};
 they can be described as
 the $S\to\infty$ limit of (\ref{lowertube}).

In Section \ref{kquarksec} we then extract from these D5-branes the value of the drag force
exerted by the $\cN=4$ SYM plasma on their dual $k$-quark bound states. Our result is given
in (\ref{fx}), which contains a non-trivial $k$-dependence displayed in
Fig.~\ref{thetakfig}. We verify that for $k=1$ and large $N$ our result matches the drag
force experienced by the trailing string of \cite{hkkky,gubser}, but (\ref{fx}) contains in
addition an infinite series of $1/N$ corrections which in analogy with \cite{df} should
codify self-interactions of the string, arising from worldsheets with an arbitrary number of
handles.

{}From the gauge theory viewpoint, a totally antisymmetric bound state of $(N-k)$ quarks
should be equivalent to a totally antisymmetric bound state of $k$ antiquarks, which by the
symmetry between quarks and antiquarks leads to the prediction that the force experienced by
a system of $k$ quarks must be the same as the one felt by $N-k$ quarks, precisely as  we
find in (\ref{fx}). This implies in particular that the force vanishes for the case of the
baryon, $k=N$, which seems natural given the no-drag result for the meson
\cite{sonnenschein,liu2,cgg}.

We also notice an intriguing connection between our drag force result (\ref{fx}) for a
$k$-quark in $\cN=4$ four-dimensional SYM and the tension $\sigma^{(k)}$ that governs the
area law for spatial Wilson loops in the same theory (or, equivalently, the tension of a
chromoelectric $k$-string in the \emph{three}-dimensional non-supersymmetric
\emph{confining} theory associated with doubly-Wick-rotated D3-branes compactified on a
small thermal circle \cite{cgst}). The precise relation between the two quantities is given
in (\ref{cgsttension}), which appeared previously in the proposal of \cite{sin2} relating
parton energy loss to magnetic confinement.  We observe that (\ref{cgsttension}) also
correctly reproduces some of the previously computed drag forces \cite{cacg1,talavera}, but
not others \cite{cacg2,mtw,ntw}, so as we discuss around (\ref{stringtension}), it would be
interesting to establish whether some type of generalization holds in a more general
context.

 In Section \ref{gluonsec} we finally proceed to the case of a gluon, which the
dictionary of \cite{gp} translates into a system of two D5-branes that respectively carry
$k=N-1$ and $k=1$ units of electric flux.  This system is described at tree-level by the
non-Abelian action (\ref{ND5}) plus an infinite series of corrections involving further
commutators and derivatives of the fields. In the $N\to\infty$ limit the two D5-branes as
expected decouple and their total energy loss rate is simply the sum of the rates of the
individual branes, which is given by our previous result (\ref{fx}). The conclusion is that,
at large $N$, the dissipative force experienced by a heavy gluon is as indicated in
(\ref{fxgluon}) double the size of the one felt by a heavy quark, precisely as group theory
predicts on the gauge theory side. The computation of $1/N$ corrections to this result would
require knowledge of the action for the two-D5-brane system at the level of the annulus and
beyond.

All the color sources considered in this paper are for simplicity taken to be infinitely
heavy, just like the external quark considered originally in \cite{gubser}. It should
however be possible to treat the case of a gluon/baryon/$k$-quark with finite mass by
introducing D7-branes \cite{kk}, in parallel with what was done for the single quark in
\cite{hkkky}. Estimates of the thermal mass of the gluon in the real-world sQGP may be found
for instance in \cite{gluonmass}.

\section{D5-brane Embedding}
\label{d5sec}

According to the prescription of \cite{yamaguchi,gp}, in order to study a bound state of
quarks in the totally antisymmetric representation of  $SU(N)$ moving through a thermal
$\cN=4$ SYM plasma, we must consider a D5-brane with electric flux that lives in the
(Schwarzschild-AdS)$_5\times\bS^5$ geometry
\begin{eqnarray}\label{metric}
ds^2&=&{1\over\sqrt{H}}\left(
-hdt^2+d\vec{x}^2\right)+{\sqrt{H}\over h}dr^2+R^2 d\Omega_5^2~, \\
H&=&{R^4\over r^4}~,\qquad  h=1-\frac{r_H^{4}}{r^4}~, \qquad R^4=4\pi N\gs\ls^4~, \nonumber
\end{eqnarray}
with self-dual Ramond-Ramond field strength
\begin{equation}\label{fiveform}
G_{(5)}\equiv dC_{(4)}=4R^4\left(\mbox{vol}(\bS^5) d\theta_1\wedge\ldots\wedge d\theta_5
-{r^3\over R^8}  dt\wedge \ldots\wedge dx_3\wedge dr\right),
\end{equation}
where $\mbox{vol}(\bS^5)\equiv\sin^4\theta_1
\mbox{vol}(\bS^4)\equiv\sin^4\theta_1\sin^3\theta_2\sin^2\theta_3\sin\theta_4$.
The D5-brane couples to these fields through the Born-Infeld plus
Chern-Simons action
\begin{eqnarray}\label{d5action}
S_{D5}&=&T_{D5}\int d^6\sigma
\left[-\sqrt{-\det\left(g_{ab}+2\pi\ap F_{ab}\right)}+
\left(2\pi\ap F_{(2)}\wedge c_{(4)}\right)_{0\ldots 5}\right]~,\\
g_{ab}&\equiv&\p_a X^{\mu}\p_b X^{\nu}G_{\mu\nu}~, \qquad
c_{a_1\ldots
a_4}\,\equiv\,\p_{a_1}X^{\mu_1}\ldots\p_{a_4}X^{\mu_4}C_{\mu_1\ldots\mu_4}~.\nonumber
\end{eqnarray}

We are instructed by \cite{yamaguchi,gp} to search for embeddings where the D5-brane extends
radially in AdS$_5$, carries a radial electric field and wraps an $\bS^4\subset\bS^5$, so it
is convenient to choose the static gauge $\sigma^a=(t,r,\theta_2,\ldots,\theta_5)$, and to
denote $\theta\equiv\theta_1$. Configurations of this type were studied previously in
\cite{imamura,cgs,pr,cpr}. As in \cite{hkkky,gubser}, we require a solution where the brane
moves at constant speed along direction $x\equiv x^1$, described by a general stationary
ansatz
\begin{equation}\label{ansatz}
X(t,r)=vt+\xi(r),\qquad \theta(r), 
\qquad
X^2=X^3=0~.
\end{equation}
Under these circumstances, the Chern-Simons term in
(\ref{d5action}) will involve only the $\theta_2\ldots\theta_5$
component of the R-R gauge field. Integrating
$\p_{\theta_1}C_{\theta_2\ldots\theta_5}=G_{\theta_1\ldots\theta_5}$,
we find it to be
\begin{equation}\label{c}
C_{\theta_2\theta_3\theta_4\theta_5}=-R^4
D(\theta)\mbox{vol}(\bS^4)~,
\end{equation}
where as in \cite{cgs} we have defined
\begin{equation}\label{d}
D(\theta)\equiv -{3\over 2}\theta+{3\over 2}\sin\theta\cos\theta+\sin^3\theta\cos\theta~.
\end{equation}
Notice that the value of the integration constant in this
expression figures in the generalized Wilson loop
$\int_{\bS^4}C_{(4)}\propto D(\theta)$, and consequently affects
the physics. Just as in \cite{pr,cpr}, we have set it equal to
zero to ensure that the solutions to follow have a consistent
physical interpretation.\footnote{The only other consistent choice
for this constant turns out to be $3\pi/2$, as in \cite{hartnoll},
and leads to equivalent physics. \label{constantfoot}}

Plugging (\ref{ansatz}) and (\ref{c}) into the action
(\ref{d5action}), the Born-Infeld and Chern-Simons terms are both
found to be proportional to $\mbox{vol}(\bS^4)$, whose integral
yields the $\bS^4$ volume $\Omega_4=8\pi^2/3$. Making use of
$T_{D5}=1/(2\pi)^5\gs\ls^6$ and the definition of $R$ in
(\ref{metric}), we are then left with
\begin{equation}\label{d5actionreduced}
S_{D5}={N\over 3\pi^2\ap}\int dt dr \left[-\sin^4\theta
\sqrt{-g^{(F1)}-(2\pi\ap F_{tr})^2}-2\pi\ap F_{tr} D\right]~,
\end{equation}
where
\begin{equation}\label{gstring}
g^{(F1)}\equiv (G_{tt}+v^2 G_{xx})(G_{rr}+\theta'^2
G_{\theta\theta})+G_{tt}G_{xx}X'^2
\end{equation}
is the determinant of the metric that would have been induced on
the worldsheet of a \emph{string} embedded in the geometry
(\ref{metric}) according to (\ref{ansatz}), and primes denote
derivatives with respect to $r$.

The equations of motion for $A_t$ and $A_r$ imply that the
momentum conjugate to $A_r$ is a constant, which the flux
quantization condition \cite{ck} requires to be an integer:
\begin{equation}\label{pia}
\Pi^t_{A_r}\equiv {\p\cL_{D5}\over\p(\p_t A_r)}={2N\over
3\pi}\left({E\sin^4\theta\over\sqrt{-g^{(F1)}-E^2}}-D\right)=k\in\bZ~,
\end{equation}
where we have set $E\equiv 2\pi\ap F_{tr}$. As is familiar from the Born-Infeld string
context \cite{cm}, the integer $k$ measures the fundamental string charge carried by the
D5-brane: $\p\cL_{D5}/\p B_{tr}=k/2\pi\ap$. It is worth emphasizing, however, that according
to (\ref{pia}) this charge receives a contribution not only from the electric field $E$ that
is relevant already in a trivial background, but also from the R-R four-form (\ref{c}). For
a given value of $k$, the equation of motion for $\theta$ allows a solution where the
D5-brane sits at a specific constant polar angle
\begin{equation}\label{thetak}
\theta=\Theta_k~,\qquad \pi {k\over N}=\Theta_k-\sin\Theta_k\cos\Theta_k~,
\end{equation}
 as long as the electric field is given by
\begin{equation}\label{esol}
E=\pm\cos\Theta_k\sqrt{-g^{(F1)}}~,
\end{equation}
with the sign chosen to coincide with that of $\sin\Theta_k$.

Notice from (\ref{thetak}) that $\Theta_{-k}=-\Theta_k$, which in (\ref{esol}) implies that
$E(-k)=-E(k)$. This encodes the expected relation between the configurations with $k$
strings and $k$ antistrings (i.e., strings with opposite orientation), which are
respectively dual to $k$ quarks and $k$ antiquarks: the corresponding D5-branes have exactly
the same embedding but opposite orientation, and carry electric fields which are equal in
magnitude but opposite in sign. There is also a close connection between the systems with
$k$ and $N-k$ strings/quarks: it follows from (\ref{thetak}) that
$\Theta_{N-k}=\pi-\Theta_{k}$, meaning that the corresponding D5-branes wrap four-spheres
located at the same angle $\Theta_k$, but measured respectively from the north and south
pole of the $\bS^5$,\footnote{Choosing the integration constant in (\ref{d}) as in the
previous footnote would simply have exchanged the role of the north and south
poles.\label{polefoot}} and (\ref{esol}) then implies that $E(N-k)=-E(k)$. The gauge theory
interpretation of this symmetry will become clear below.

In the extremal case $r_H=0$, Lorentz invariance of (\ref{metric}) along the boundary
directions guarantees that the configuration (\ref{ansatz}) where the D5-brane moves along
$x$ with speed $v$ is just a boosted version of the static solution, so we must have
$\xi(r)=0$ and therefore $X'=0$, i.e., the brane remains upright. The authors of
\cite{yamaguchi,gp} worked precisely in this case, and identified the constant-polar-angle
solution (\ref{thetak})-(\ref{esol}), first obtained in \cite{pr}, as the AdS counterpart of
a bound state, in zero-temperature $\cN=4$ SYM, of $k$ external quarks in the totally
antisymmetric representation, moving at constant velocity $v$. Their identification was
principally based on the fact that, just like a purely radial fundamental string, the above
D5-brane embeddings preserve the same fermionic and bosonic symmetries as the required gauge
theory sources: they are half-BPS and invariant under a $SO(2,1)\times SO(3)\simeq
SU(1,1)\times SU(2)$ subgroup (generated by $P_0,K_0,D$ and $J_1,J_2,J_3$) of the $SO(4,2)$
conformal group, as well as a $SO(5)$ subgroup of the $SO(6)\simeq SU(4)$ R-symmetry
group--- altogether, they are $Osp(4^*|4)$ invariant. A puzzling feature of these
configurations, emphasized already in \cite{pr}, is that their energy is not proportional to
their charge. We will come back to this puzzle below.

Interestingly, in work prior to \cite{pr}, a seemingly different family of D5-brane
configurations that are also half-BPS was obtained and proposed to describe a $k$-quark
bound state \cite{cgs}. These are also solutions of the equations that follow from the
extremal version of (\ref{d5actionreduced}), but have a non-trivial profile $\theta(r)$,
whose inverse was determined analytically in \cite{cgs},
\begin{equation}\label{lowertube}
r(\theta)={S\over\sin\theta}\left[(\pi k/N-\theta+\sin\theta\cos\theta)\over \pi
k/N\right]^{1/3}~,
\end{equation}
with $S$ an arbitrary constant, and carry a corresponding electric field as dictated by
(\ref{pia}). These solutions cover the angular range $0\le\theta\le\Theta_k$ (where the
expression inside the brackets is positive) and have $r$ decreasing monotonically from
$\infty$ at $\theta=0$ to $0$ at $\theta=\Theta_k$. They describe a bundle of $k$
fundamental strings that extend radially from the boundary at $r\to\infty$ to the horizon at
$r=0$, pointing always towards the north pole of $\bS^5$, $\theta=0$.

The embeddings (\ref{lowertube}) were shown to be BPS from the point of view of
supersymmetry preservation in \cite{imamura,cgmvp}, and from the perspective of energy
minimization in \cite{cgs,cgmvp,grst}. They leave 16 supersymmetries unbroken, and have an
energy equal to that of $k$ fundamental strings, supporting their identification as
threshold bound states of $k$ quarks in $\cN=4$ SYM \cite{cgs}. This holds for any value of
the scale factor $S$, which is therefore a modulus in the space of equal-energy
configurations. Through the standard UV-IR connection \cite{uvir}, on the gauge theory side
$L\equiv R^2/S$ is a measure of the spatial extent of the $k$-quark bound state, as seen
most explicitly in \cite{cg} for the case $k=N$. The existence of solutions with arbitrary
size is of course a consequence of the conformal invariance of $\cN=4$ SYM.

Given that both  (\ref{thetak})-(\ref{esol}) and the one-parameter family (\ref{lowertube})
are half-BPS D5-brane configurations claimed to be dual to $k$-quark bound states, it is
natural to wonder whether they are related in some way. An important clue in this regard
comes from the fact that, although the fermionic symmetries they preserve are the same,
their bosonic symmetries do not quite coincide. For general scale factor $S$, the embedding
(\ref{lowertube}) is, just like (\ref{thetak})-(\ref{esol}), invariant under $SO(5)\subset
SO(6)$ and $SO(3)\subset SO(4,2)$, but not under $SO(2,1)\subset SO(4,2)$, because the
dilatation generator $D$ sends $(t,\vec{x},r)\to (\lambda t,\lambda \vec{x},r/\lambda)$ and
consequently rescales the value of $S$.

The only exception to this is the case with $S\to\infty$ (corresponding to a gauge theory
source with size $L\to 0$), which \emph{is} invariant under dilatations, and, altogether,
under precisely the \emph{same} $Osp(4^*|4)$ as the constant-angle embedding found in
\cite{pr}. The reason for this agreement is plain to see: in the $S\to\infty$ limit, the
portion of (\ref{lowertube}) lying near $r=0$ and $\theta=\Theta_k$ is blown up to coincide
\emph{exactly} with (\ref{thetak})-(\ref{esol})! The constant-angle D5 embedding obtained in
\cite{pr} is in this way understood to be a particular member of the one-parameter family of
configurations found earlier in \cite{cgs}. Moreover, the solution of \cite{pr} is seen to
be incomplete, because it is missing the portion of (\ref{lowertube}) that covers the
angular range $0\le\theta<\Theta_k$, which is pushed by the $S\to\infty$ limit all the way
out to the AdS boundary $r\to\infty$. To put it in more graphic language, each of the polar
plots of (\ref{lowertube}) shown in Fig.~1 of \cite{cgs} can be described as a small `cone'
joined onto a long `tube', and the solution (\ref{thetak})-(\ref{esol}) is obtained, after
an infinite rescaling, by keeping the now infinitely large cone but ignoring the tube, which
now lies at the boundary and can be thought of as an end cap for the cone.

This last observation resolves the energy-charge puzzle that appeared in \cite{pr}: once the
end cap is included, the total energy of the configuration \emph{is} found to be correctly
proportional to its total charge, as demonstrated in \cite{cgs}.\footnote{Notice that the
discussion in \cite{cgs} of a `missing charge' for the cases with $k\neq 0,N/2,N$ was
misguided: the relevant charge is in all cases given by $k$ as in (\ref{pia}), and
\emph{not} by the total R-R flux intercepted by the D5-brane.} This clarification is
particularly important in the case of the baryon, $k=N$, where we find $\Theta_N=\pi$
meaning that the cone (\ref{thetak})-(\ref{esol}) has collapsed to become a tensionless
string, but the end cap now consists of a D5-brane fully wrapped around the $\bS^5$ at
$r\to\infty$ and carries an energy equal to that of $N$ quarks.

The presence of the end cap also plays a role in explaining how the D5 embedding codifies
the $SU(4)$ quantum numbers of the quarks. Indeed, unlike the cone (\ref{thetak}), which
lies in the northern or southern hemisphere of $\bS^5$ depending on whether $k<N/2$ or
$k>N/2$, the end cap included in the full solution (\ref{lowertube}) always closes off at
the \emph{north} pole, correctly reflecting the fact that the D5-brane in question describes
a bundle of strings (or antistrings) pointing towards $\theta=0$, and is consequently dual
to $k$ quarks (or antiquarks) which source only the specific SYM scalar field that defines
the polar direction of the internal space. These solutions were called `lower tubes' in
\cite{cgs}. That same work constructed as well `upper tube' solutions that always close off
at the \emph{south} pole and describe a bundle of antistrings (or strings)  pointing towards
$\theta=\pi$.\footnote{These are comparable to the solutions of \cite{pr} that result from
the alternative choice of integration constant described in footnotes \ref{constantfoot} and
\ref{polefoot}.} The embedding of this type that, based on energetics, corresponds to a
bound state of $k$ antistrings (or strings), is found to consist of a cone at
$\theta=\Theta_{N-k}$ and an end cap that covers the angular range
$\Theta_{N-k}<\theta\le\pi$. By symmetry, the angle that this `upper' cone makes with the
south pole, $\pi-\Theta_{N-k}$, must coincide with the angle   $\Theta_k$ that the `lower'
cone describing $k$ antistrings (or strings) makes with the north pole. This explains the
previously mentioned connection $\Theta_{N-k}=\pi-\Theta_k$, which we now recognize as the
AdS counterpart of the gauge-theoretic statement that a totally-antisymmetric bound state of
$N-k$ quarks with a given $SU(4)$ charge is equivalent to a  totally-antisymmetric bound
state of $k$ \emph{anti}quarks with the \emph{opposite} $SU(4)$ charge.

In the last few paragraphs we have focused on the extremal background, which was the arena
of \cite{cgs,pr,yamaguchi,gp}. Since in this paper we wish to study the effects of a thermal
$\cN=4$ SYM plasma, we are mostly concerned with D5-brane embeddings in the
\emph{non}-extremal geometry (\ref{metric}) with $r_H>0$. As we have seen above, in that
case the constant-angle embedding (\ref{thetak})-(\ref{esol}) is still a solution, as was
first proven in \cite{hartnoll}. The configurations (\ref{lowertube}), on the other hand, do
not admit a straightforward generalization to the non-extremal case \cite{cgst,gst}. An
important point is that the presence of the black hole in AdS breaks supersymmetry and
conformal invariance, so  $S$ can no longer be a modulus, and we would \emph{not} in fact
expect to find a one-parameter family of solutions directly generalizing (\ref{lowertube}).
But given that the cone (\ref{thetak}) \emph{is} a solution,  it should be the case that the
$S\to\infty$ limit of (\ref{lowertube}) in which it is contained is also a solution. To
verify this one need only show that the portion of the D5-brane which is contained in the
full embedding (\ref{lowertube}) and not in (\ref{thetak}), i.e., the end cap, solves the
non-extremal equation of motion, but this is evidently true because this portion lies
entirely at $r\to\infty$, where the terms in the equation arising from the non-extremality
are subleading. This is then another important property that distinguishes the $S\to\infty$
configurations from its $S<\infty$ siblings: among the $k$-quark bound states found in
\cite{cgs}, it is the only one that admits a direct generalization away from the extremal
case. Clearly it is this embedding alone that will provide us with the pointlike ($L=0$)
source that we wish to drag through the thermal plasma, and for this reason in the sections
to follow we will concentrate exclusively on it.

\section{Drag Force on $k$-Quarks and Baryons}
\label{kquarksec}

In the previous section we have learned that a bound state of $k$ quarks in the totally
antisymmetric representation moving through a thermal $\cN=4$ SYM plasma is described in
dual language by a D5-brane on (Schwarzschild-AdS)$_5\times\bS^5$ whose $\theta(r)$ profile
is given by the $S\to\infty$ limit of (\ref{lowertube}), which consists of two parts: the
`cone' (\ref{thetak})-(\ref{esol}), which extends from the horizon $r=r_H$ to the boundary
$r\to\infty$, and an `end cap' located at the boundary and covering the angular range
$0\le\theta<\Theta_k$. As we have seen, the inclusion of this end cap is important to
reproduce the correct energy and $SU(4)$ charge of the system of $k$ quarks, and in fact the
\emph{same} cone, with a $\Theta_k<\theta\le\pi$ end cap, is dual to a bound state of $N-k$
antiquarks. We would now like to determine the force $F_x$ with which this D5-brane must be
pulled to maintain its velocity, which is mapped by AdS/CFT into the drag force experienced
by the $k$-quark (or $(N-k)$-antiquark). As in \cite{hkkky,gubser}, this force will be
non-zero only if the moving D5-brane leans back, i.e., if the function $\xi(r)$ describing
the profile (\ref{ansatz}) of the D5-brane in the direction of motion is non-trivial. Since
the end cap lies only at the boundary, for the purpose of determining this profile we need
only concentrate on the conical portion of the solution.

For the cone (\ref{thetak})-(\ref{esol}), the equation of motion for $X(r,t)$ that follows
from (\ref{d5actionreduced}) is easily seen to be the same as the one that would be obtained
from the Nambu-Goto Lagrangian $-\sqrt{-g^{(F1)}}$, which describes a \emph{string} embedded
in (Schwarzschild-AdS)$_5$ according to (\ref{ansatz}), with fixed
$\theta_1,\ldots,\theta_5$. This is just a particular case of the result obtained in
\cite{hartnoll}, where it was shown that, given \emph{any} solution to the equations of
motion for a string that reaches the boundary of an asymptotically AdS$_5$ spacetime $M$,
one can obtain a one-parameter family of valid D5-brane embeddings on $M\times\bS^5$ (where
the sphere is threaded by $N$ units of R-R flux), simply by taking the D5 to carry $k$ units
of electric flux and wrap the $\bS^4\subset\bS^5$ at polar angle $\theta=\Theta_k$. Based on
this general result, the drag force we obtain below in the context of $\cN=4$ SYM should
also be relevant for \emph{any} thermal gauge theory with a dual description in terms of
Type IIB string theory on $M\times\bS^5$.

Because the reduced dynamics coincides with that of a string, the
situation is exactly the same as the one examined in
\cite{hkkky,gubser}: for stationary configurations of the type
(\ref{ansatz}), the equation of motion for $X(r,t)$ implies that
the conjugate momentum density
\begin{equation}\label{pix}
\Pi^r_x\equiv {\p\cL_{D5}\over\p X'}={2N\sin^3\Theta_k\over
3\pi}{1\over
2\pi\ap}\left({G_{tt}G_{xx}X'\over\sqrt{-g^{(F1)}}}\right)
\end{equation}
is a constant, and there is a critical radius $r_v\equiv r_H/(1-v^2)^{1/4}$ below which the
solution is found to be real only if the factor within parentheses takes the precise value
$-(r_H/R)^2(v/\sqrt{1-v^2})$. This then fixes the value of the momentum density (\ref{pix}),
which measures the rate at which $x$-momentum flows radially along the D5-brane, away from
the boundary and towards the horizon. To keep it moving at constant velocity $v$ along
direction $x$, an external agent must supply to the D5-brane precisely this amount of linear
momentum at the boundary, i.e., it must exert a force of magnitude (\ref{pix}).

According to the AdS/CFT dictionary established in \cite{yamaguchi,gp}, the force exerted on
the D5-brane translates into the drag force experienced by a system of $k$ quarks (or $N-k$
antiquarks) in the totally antisymmetric representation that ploughs through a thermal
$\cN=4$ SYM plasma. Through use of the relations $T=r_H/\pi R^2$ and $R^4=g_{YM}^2 N\ls^2$,
this dissipative force may be written in the final form
\begin{equation}\label{fx}
F_x^{(k)}\equiv {dp_x\over dt}={2N\over 3\pi}\sin^3\Theta_k\left(-{\pi\over 2}\sqrt{g_{YM}^2
N}T^2 {v\over\sqrt{1-v^2}}\right)~,
\end{equation}
which is our main result in this section. The momentum flow (\ref{fx}) from the moving
$k$-quark to the plasma implies a corresponding energy loss rate $dE/dt=vdE/dx= vF_x^{(k)}$
(which can also be inferred directly from the D5-brane momentum density $\Pi^r_t$).

\vspace{1.0cm}

\begin{figure}[tbh]
\centering \setlength{\unitlength}{1cm}
\includegraphics[width=8cm,height=4cm]{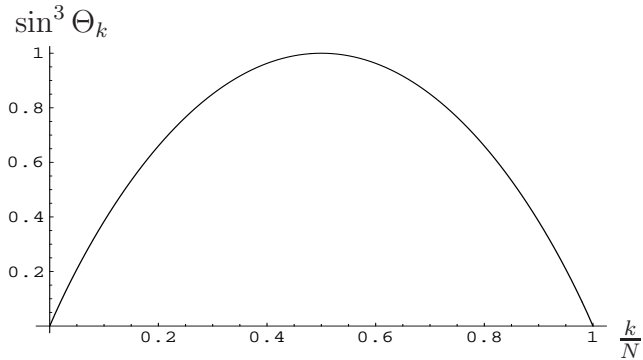}
 \begin{picture}(0,0)
\put(-8,4.2){$\sin^3{\Theta_k}$} \put(0,0){$\frac{k}{N}$}
\end{picture}
\vskip.5cm \caption{$k$-dependence of the drag force experienced by a
$k$-quark.}\label{thetakfig}
\end{figure}

It is interesting that the $k$-dependence is contained entirely in the $\sin^3\Theta_k$
prefactor, whose behavior is shown in Fig.~\ref{thetakfig}, and is such that
$F_x^{(N-k)}=F_x^{(k)}$, exhibiting the expected symmetry between quarks and antiquarks.
This implies in particular that the bound state of $N$ quarks, the baryon, must feel the
same force as a system with no quarks at all, which is to say that $F_x^{(N)}=0$. We thus
learn that, at this level of approximation, the baryon can traverse the plasma without
suffering any energy loss. This is the same conclusion as was reached for the meson in
\cite{sonnenschein,liu2,cgg}. The reason in both cases is the same: neither of these
color-neutral sources are able to set up the long-range gluonic fields\footnote{The
chromoelectric field produced by the baryon has been determined explicitly in \cite{cg} at
zero temperature.} that are responsible for carrying energy away from the $k$-quarks with
$0<k<N$, as has been studied in detail in \cite{gubseretal1,gubseretal2}. In fact, for $k=N$
the angle (\ref{thetak}) becomes $\theta=\pi$, meaning that the `conical' part of the
D5-brane has closed off and become tensionless. Since it is this portion that codifies the
gluonic fields set up by the $N$-quark system, we conclude that our pointlike baryon
produces no field at all. It is in effect described solely by the D5 `end cap' at the AdS
boundary, which in this case is completely wrapped around $\bS^5$, and is able to move
unimpeded because, lying at $r\to\infty$, it is unaffected by the presence of the horizon.

It is instructive to consider the case of a single quark, $k=1$ (or a single antiquark,
$k=N-1$), where the D5-brane lies at a polar angle $\Theta_1$ (or $\pi-\Theta_1$) which
according to (\ref{thetak}) is given by $\pi/N=\Theta_1-\sin\Theta_1\cos\Theta_1$. For $N\gg
1$, this implies that $\sin^3\Theta_1 \simeq\Theta_1^3\simeq 3\pi/2N$, so as required the
drag force $F_x^{(1)}|_{N\to\infty}$ agrees with the one obtained in \cite{hkkky,gubser}
(the factor inside the parentheses in (\ref{fx})), whose computation was carried out at
string tree-level, and therefore applies in the large $N$ limit. Based on the results of
\cite{df}, it is tempting to speculate that the finite $N$ corrections to this result that
can be deduced from the D5-brane drag force (\ref{fx}) encode higher-order corrections that
arise from self-interactions of the string. More generally, if we hold $k$ fixed in the
large $N$ limit, it follows from (\ref{fx}) that
\begin{equation}
F_x^{(k)}=k\left(1-{3\over 10}\left({3\pi k\over 2 N}\right)^{2/3}+\ldots\right)
F_x^{(1)}|_{N\to\infty}~,
\end{equation}
and we would expect the second and higher terms to capture the effect of the interactions of
the $k$ quarks among themselves. Notice that the first correction is negative, meaning that,
at finite $g_{YM}$, the $k$-quark loses energy at a somewhat smaller rate than $k$ unbound
quarks. It should of course be borne in mind that, since unlike \cite{df} we are not working
in a supersymmetric context, there might be additional finite-$N$ contributions arising from
string-loop corrections to the background (\ref{metric}).

The $\sin^3\Theta_k$ prefactor seen in our drag force result (\ref{fx}) also appeared in the
Wilson loop calculation of \cite{hartnoll}. This is evidently not surprising from the string
theory perspective, because both calculations use the same D5-brane embedding, the sole
difference being that the value of the Wilson loop receives a contribution not only from the
action (\ref{d5actionreduced}) but also from various boundary terms that are responsible for
yielding a finite answer \cite{dgo,df,hpk2}. The agreement is even more unsurprising from
the gauge theory viewpoint, because both quantities refer to the same type of external
source in the same thermal gauge theory.

What we do find interesting is that the very same factor $\sin^3\Theta_k$ showed up as well
in \cite{cgst}, which dealt with a problem that from the field theory perspective would
appear to be completely different: the computation of flux tube tensions $\sigma^{(k)}$ in
the confining three-dimensional gauge theory that is obtained by dimensionally reducing
Wick-rotated D3-branes on a supersymmetry-violating circle (in parallel with the D4-brane
discussion of \cite{wittenthermal}). This problem was studied in \cite{cgst} by splitting a
baryon into two groups of $k$ and $N-k$ quarks (or equivalently, $k$ quarks and $k$
antiquarks), thereby producing a long chromoelectric flux tube, a QCD $k$-string, in
between. The dual description for this split baryon involves a D5-brane (as inferred from
the picture developed for $\cN=4$ SYM in \cite{wittenbaryon,imamura,cgs}), and the portion
of this brane that is dual to the flux tube turns out to lie precisely at the polar angle
(\ref{thetak}), which explains, from the string theory side, the agreement with our result.

The precise relation between the quantities computed here and in \cite{cgst} is
\begin{equation} \label{cgsttension}
F_x^{(k)}=-\sigma^{(k)}{v\over\sqrt{1-v^2}}~,
\end{equation}
where it is seen that the force and the tension have identical dependence on $k$, $N$, the
't~Hooft coupling and the temperature. Using a fundamental string instead of a D5-brane, the
authors of \cite{sin2} found this  same relation to hold in the particular case $k=1$,
$N\to\infty$. They interpreted $\sigma^{(k)}$ directly in the four-dimensional SYM theory as
the tension controlling the area law for \emph{spatial} Wilson loops \cite{wittenthermal},
and used the fact that such loops codify the magnetic interaction between two current wires
to argue that the agreement (\ref{cgsttension}) reflects the magnetic origin of the drag
force. We will return to this point below.

 The computation of the $k$-string tension in $d$-dimensions is an important problem in
the study of confining gauge theories (see, e.g., \cite{shifman} for recent reviews, and the
works \cite{fluxtubes} for related AdS/CFT calculations), and it would be worth exploring
whether the connection obtained here and in \cite{sin2} with a drag force calculation in
$d+1$ dimensions holds more generally. For instance, the simple $(k/N)(1-k/N)$ scaling of
the flux tube tension found in \cite{cgst} for the four-dimensional Yang-Mills theory
obtained from D4-branes \cite{wittenthermal} suggests an identical scaling for the rate of
parton energy loss in the corresponding five-dimensional thermal plasma. Notice that the
former quantity is static, whereas the latter is dynamical, so if one could establish a
relation of the type (\ref{cgsttension}) in a more general setting, it would in particular
become possible to access information on energy loss from flux tube calculations on the
lattice.

A formula like (\ref{cgsttension}) would also greatly facilitate the AdS/CFT computation of
parton energy loss in other theories, because on the geometry side the value of
$\sigma^{(k)}$ is simply given by the tension of a string/brane that is localized in time
and lies at the stationary limit surface $r=r_s$ where $G_{tt}=0$. In the case of a source
in the fundamental representation, which is modelled with a fundamental string, this amounts
to $\sigma^{(1)}=G_{xx}(r_s)/2\pi\ap$ (with $G_{\mu\nu}$ the string frame metric). We have
checked that when inserted in (\ref{cgsttension}), this correctly reproduces the drag force
computed in some cases \cite{cacg1,talavera}, but not in others \cite{cacg2,mtw,ntw}, which
generally involve non-trivial background fields other than the metric. In particular, the
$v$-dependence of the actual drag force can be much more complicated than $v/\sqrt{1-v^2}$,
so the desired generalization of (\ref{cgsttension}), if it exists at all, should involve a
prescription for correctly determining this dependence.

Given that in the drag force calculations a central role is played not by the stationary
limit radius $r_s$ but by the velocity-dependent radius $r_v$, it is natural to expect a
relation between $F_x$ and the tension of a spacelike string (or brane) localized at the
latter radius. And indeed, working on an arbitrary background as in \cite{herzog,cacg2} one
can prove that in all cases
\begin{equation}\label{stringtension}
F_x=-\tilde{\sigma}{v\over\sqrt{1+v^2}}~,
\end{equation}
where $\tilde{\sigma}=\sqrt{1+v^2}G_{xx}(r_v)/2\pi\ap$ is the tension of a string worldsheet
extended along $\tilde{x},x_2$ and localized at $\tilde{t}=\mbox{const.}$, $r=r_v$, with
$(\tilde{t},\tilde{x})$ the coordinates in the $q$-$\bar{q}$ rest frame and $r_v$ the radius
where $G_{\tilde{t}\tilde{t}}=0$. This last condition shows that $r_v$ marks the location of
the stationary limit surface of the black brane \emph{as seen by an observer at rest in the
$(\tilde{t},\tilde{x})$ frame} (and so in particular agrees with $r_s$ when $v=0$). All of
this strongly suggests that (\ref{stringtension}) is the desired generalization of the
force-tension relation (\ref{cgsttension}). But for this to be useful, a simple
gauge-theoretic interpretation should be found for the tension $\tilde{\sigma}$, which has
thus far been defined only on the AdS side. The natural guess that it controls the area law
for a spatial Wilson loop extended along $\tilde{x},x_2$ turns out to be incorrect, because
the $\cap$-shaped string that would be employed on the AdS side to compute such a loop
descends beyond $r_v$ and all the way down to $r_s$ as the loop becomes large. This casts
some doubt on the proposal of \cite{sin2}, because (\ref{cgsttension}) appears to work only
in those cases where the background is such that $\sigma$ happens to be proportional to
$\tilde{\sigma}$.

In any event, it should be noted that the observation made in \cite{sin2} and the present
paper that the information on the rate of parton energy loss can (at least in some cases) be
extracted from a \emph{spatial} Wilson loop is reminiscent of the proposal of \cite{liu} for
computing the jet-quenching parameter using a \emph{lightlike} Wilson loop, and in fact
lends credence to the idea that the latter is to be regarded as a limit of a
\emph{spacelike} loop traced by a pointlike source whose velocity approaches the speed of
light \emph{from above} \cite{cgg}. Notice, however, that the loops discussed here and in
\cite{liu} explore different regimes: whereas the magnetic area law ($E\propto L$) is found
to hold for wide loops ($LT\gg 1$), the `dipole approximation' ($E\propto L^2$) that
inspired \cite{liu} is valid only for narrow loops ($LT\ll 1$). It might prove worthwhile to
investigate in detail the possible extrapolation between these two regimes, hopefully
shedding light on the meaning of (\ref{stringtension}) in the process.

It would also be interesting to extend our drag force calculation to the case of external
sources in representations other than the totally antisymmetric one. The simplest example
would be a system of $k$ quarks in the totally symmetric representation, which according to
the dictionary of \cite{gp} is dual to a D3-brane that carries $k$ units of fundamental
string charge. The equations that determine the relevant D3-brane embedding are much more
complicated than the ones for the D5-brane, so regrettably we must leave their analysis to
future work.

\section{Drag Force on a Heavy Gluon}
\label{gluonsec}

According to \cite{gp}, in the same way we represent $k$ quarks in the totally antisymmetric
representation as a D5-brane with $k$ units of fundamental string charge, we can think of a
gluon (a source in the adjoint representation) as two parallel D5-branes, carrying $k=1$ and
$k=N-1$ units of fundamental string charge respectively. The dynamics of this system is
described by some $U(2)$ action whose full form is not yet known. To date, the most complete
proposal is that of \cite{Myers} (see also \cite{tvr}),
\begin{eqnarray}\label{nA}
S_{Dp}&=&-T_p\int{d^{p+1}\sigma \tr\left(e^{-\phi}\sqrt{-\det\left(P\left[E_{\mu\nu}+E_{\mu
i}(Q^{-1}-\delta)^{ij}E_{j\nu} \right]+2\pi{\alpha}' F_{\mu\nu}  \right)\det(Q^i_j) }
\right)
}\nonumber\\
{}&{}& + T_p\int{\tr\left(P\left[e^{i2\pi{\alpha}' \textrm{i}_\Phi \textrm{i}_\Phi}(\sum
C^{(n)}e^B) \right]e^{2\pi{\alpha}'  F}  \right) }
\end{eqnarray}
where
\begin{equation}
E_{\mu\nu}=G_{\mu\nu}+B_{\mu\nu}, \qquad Q^i_j=\delta^i_j+i 2\pi\ap\left[
\Phi^i,\Phi^k\right]E_{kj},
\end{equation}
$P[E]$ is the general formula for the non-Abelian pullback and $\textrm{i}_\Phi$ denotes the
interior product by $\Phi_i$ regarded as a vector in the transverse space. The gauge field
is now non-Abelian and the scalars ${\Phi}^i$ belong to the adjoint $U(2)$ representation.
As first noticed by  \cite{Tseytlin,Hull}, all derivatives in the action are replaced by
covariant derivatives. Furthermore, all the bulk fields are functionals of the non-Abelian
scalars. As in the Abelian case, these fields are interpreted in terms of a Taylor
expansion, however, the transverse displacements are now matrix-valued, so the action would
be given by a non-Abelian Taylor expansion \cite{Myers}.

Finally, the gauge trace in (\ref{nA}) is meant to be implemented as was first proposed in
\cite{Tseytlin}, taking the totally symmetrized product of all non-Abelian expressions of
the form $F_{\mu\nu}$, $D_{\mu}\Phi^i$ and $\left[\Phi^i,\Phi^j\right]$. This prescription
correctly yields the $F^2$ and $F^4$ interactions, but seems to require modifications at
order $F^6$ and higher \cite{Bain}. This shortcoming is related to the fact that the
prescription of dropping all interactions involving derivatives of the field strength, as
one does in the Abelian Born-Infeld action, is ambiguous in the non-Abelian theory because
of the commutators that involve the gauge fields.

In our particular case, where $\phi=0$, $B_{\mu\nu}=0$ and the spacetime metric
(\ref{metric}) is diagonal, (\ref{nA}) reduces to:
\begin{equation}\label{ND5}
S_{D5}=-T_{D5}\int{d^{6}\sigma \tr \left[\sqrt{-\det\left(P[G_{\mu\nu}]+ 2\pi{\alpha}'
F_{\mu\nu} \right)}-\left(2\pi{\alpha}'F_{(2)}\wedge P[C_{(4)}] \right)_{0\cdots 5} \right]
}
\end{equation}
with
\begin{equation}
P[G_{\mu\nu}]=G_{\mu\nu}+G_{ij}D_{\mu}X^iD_{\nu}X^j \qquad
F_{\mu\nu}=\partial_{\mu}A_{\nu}-\partial_{\nu}A_{\mu}+i\left[A_{\mu},A_{\nu} \right]
\end{equation}

Upon expanding both terms in (\ref{ND5}) to get a polynomial expression in the fields, it is
easy to show that, after taking the trace, all the off-diagonal elements ($\scriptstyle{12}$
or $\scriptstyle{21}$) of the fields always appear in pairs. It is therefore consistent with
the equations of motion to set all these elements equal to zero. Furthermore, all
interactions between diagonal terms of type $\scriptstyle{11}$ and $\scriptstyle{22}$
involve also off-diagonal elements, and consequently setting the latter to zero turns off
these interactions. So, at this order, the two D5-branes do not see one another, and we can
work directly with two copies of the Abelian equations of motion.

It is important to notice that the above result applies in fact not only for the specific
action (\ref{nA}), but also for whatever action codifies the complete tree-level D-brane
interactions. The reason is that this action arises from disk diagrams, where
the presence of a single boundary (responsible for the appearance of a
single trace) forces each $\scriptstyle{12}$ vertex to be paired with a $\scriptstyle{21}$
vertex, and allows insertion of $\scriptstyle{11}$ and $\scriptstyle{22}$ vertices in a
single diagram only in conjunction with at least one of these off-diagonal pairs.

At this point, we can use our results from Section \ref{kquarksec}. We have two decoupled
D5-branes, one with $k=1$ and the other with $k=N-1$ units of electric flux. Using equation
(\ref{fx}) we then get the net drag force on the gluon,
\begin{equation} \label{fxgluon}
F(\mbox{gluon})=F^{(1)}_x+F^{(N-1)}_x=2F^{(1)}_x=2F(\mbox{quark})~.
\end{equation}
This result is in agreement with what is expected from the gauge theory side. The basic
point is that, just from the Feynman rules for the three-point QCD vertices, one finds that
the probability for a parton to radiate a gluon is proportional to the quadratic Casimir in
the relevant representation of the $SU(N)$ group, $N$ if the parton is a gluon and
$(N^2-1)/{2N}$ if it is a quark. The same group-theory factor appears in the more
complicated calculations of parton energy loss \cite{baier}, and for this reason the
relative factor between the gluon and quark energy loss rate is expected to be
$2N^2/(N^2-1)$ \cite{guy,d'Enterria}, at least in the perturbative regime. In the large $N$
limit, this reduces to a factor of 2, just as we have found here. In the real-world $N=3$
case, the factor is not quite 2 but 2.25.

Our result should receive $1/N$ corrections, which must come from calculations at higher
order in string loops. 
Starting at the level of the annulus, the presence of multiple boundaries (which results in
multiple traces) allows simultaneous insertions of $\scriptstyle{11}$ and $\scriptstyle{22}$
vertices without the appearance of off-diagonal modes (which still come in pairs, and so can
consistently be set to zero), meaning that we can no longer treat each D5-brane
independently. These diagrams are suppressed by powers of $\gs=g_{YM}^2/4\pi$, which for
fixed $g_{YM}^2 N$ is equivalent to powers of $1/N$. It would be interesting to work out
these corrections, but it would require explicit knowledge of the higher-order form of the
non-Abelian action.

\section*{Acknowledgements}

We thank Alejandro Ayala, Eleazar Cuautle, Antonio Garc\'{\i}a, Mart\'{\i}n Kruczenski and
Guy Paic for very valuable discussions and for pointing out several useful references. We
are also grateful to Antonio Garc\'{\i}a and Dami\'an Hern\'andez for comments on the
manuscript. This work was partially supported by Mexico's National Council of Science and
Technology (CONACyT) grants CONACyT 40754-F and CONACyT SEP-2004-C01-47211, as well as by
DGAPA-UNAM grant IN104503-3.

\end{document}